\documentclass{article}

\usepackage{arxiv}

\usepackage[utf8]{inputenc} 
\usepackage[T1]{fontenc}    
\usepackage[english]{babel} 
\usepackage{hyperref}       
\usepackage{url}            
\usepackage{amsfonts}       
\usepackage{nicefrac}       
\usepackage{microtype}      
\usepackage{amsmath}        
\usepackage{graphicx}       
\usepackage{subfigure}      
\usepackage[version=4]{mhchem}  

\title{Relation between the convective field and the stationary probability distribution of chemical reaction networks}

\author{
  Lara Becker \\
  Institut für Festkörperphysik\\
  Technische Universität Darmstadt\\
  Hochschulstr. 6, 64289 Darmstadt, Germany \\
  \texttt{larabecker@fkp.tu-darmstadt.de} \\
   \And
  Marc Mendler \\
   Institut für Festkörperphysik\\
  Technische Universität Darmstadt\\
  Hochschulstr. 6, 64289 Darmstadt, Germany \\
  \texttt{marcm@fkp.tu-darmstadt.de} \\
  \And
  Barbara Drossel \\
  Institut für Festkörperphysik\\
  Technische Universität Darmstadt\\
  Hochschulstr. 6, 64289 Darmstadt, Germany \\
  \texttt{drossel@fkp.tu-darmstadt.de} \\
}

\begin{document}
\maketitle

\begin{abstract}
We investigate the relation between the stationary probability distribution of chemical reaction systems and the convective field derived from the chemical Fokker-Planck equation (CFPE) by comparing predictions of the convective field to the results of stochastic simulations based on Gillespie's algorithm. The convective field takes into account the drift term of the CFPE and the reaction bias introduced by the diffusion term. For one-dimensional systems, fixed points and bifurcations of the convective field correspond to extrema and phenomenological bifurcations of the stationary probability distribution whenever the CFPE is a good approximation to the stochastic dynamics. This provides an efficient way to calculate the effect of system size on the number and location of probability maxima and their phenomenological bifurcations in parameter space. For two-dimensional systems, we study models that have saddle-node and Hopf bifurcations in the macroscopic limit. Here, the existence of two stable fixed points of the convective field correlates either with two peaks of the stationary probability distribution, or with a peak and a shoulder. In contrast, a Hopf bifurcation that occurs in the convective field for decreasing system size is not accompanied by the  onset of a crater-shaped probability distribution; decreasing system size rather destroys craters and replaces them by local maxima. 
\end{abstract}


\section{Introduction}
On all levels of biology, systems are subject to noise, with examples ranging from the demographic stochasticity of populations in ecology to the fluctuating concentrations of proteins and mRNA transcripts in individual cells~\cite{Tsimring2014, Black2012}. Intrinsic stochasticity thus represents a general condition under which most biological systems operate, especially on the cellular level~\cite{Shahrezaei2008, Tsimring2014}. Intrinsic stochasticity can change the behavior of a system substantially compared to the dynamics of the same system in the absence of noise. Examples of this are the induction of repeated transitions between two stable states, resulting in what is called bistability~\cite{Gardiner2009}, or the creation of new stables states in the stochastic system by noise~\cite{Biancalani2014, Bishop2010}. Stochasticity has also been shown to induce quasi-oscillatory behavior for systems which do not show sustained oscillations in the macroscopic limit~\cite{Alonso2006, Boland2008, McKane2005, McKane2007, Thomas2013, Vilar2002}. 

Many systems of biological and (bio-)chemical interest can be described by (chemical) reaction networks~\cite{Angeli2009, Gunawardena2003}. But despite the importance of intrinsic stochasticity, the prediction of its effects often remain elusive. The simplest models for the dynamics of chemical reaction networks stem from dynamical systems theory. They capture system behavior in the absence of noise via sets of ordinary differential equations (ODEs) governing the time evolution of the state variables. Written down in the form
\begin{equation}
    \frac{\text{d}\vec{c}}{\text{dt}} = \vec{f}(\vec{c}),
\label{eq:DynamischesSystemIntroduction}
\end{equation}
they are called dynamical systems~\cite{Strogatz1994}. A wide range of methods is available for the analysis of such systems, such as the analysis of the topology of the vector field $\vec{f}(\vec{c})$ and the study of bifurcation diagrams as well as stability diagrams. Unfortunately, these methods cannot easily be transferred to stochastic systems as it is not per se clear how to include the effects of intrinsic noise in the vector field~$\vec{f}(\vec{c})$. A means of doing so was recently put forward by Mendler et al.~\cite{Mendler2018}, who used the so-called convective field in order to analyze how stochasticity changes the behavior obtained from the macroscopic rate equations. Based on the chemical Fokker-Planck equation (CFPE), this convective field contains in addition to $\vec{f}(\vec{c})$ a term that takes into account reaction biases introduced by the intrinsic noise, and it can be analyzed in the same way as $\vec{f}(\vec{c})$. 

The method of Mendler et al.~is based on the insight that for vanishing stationary probability currents $\vec{j}_{s}(\vec{c})$ of the CFPE, stable and completely unstable fixed points (i.e., sinks and sources) of the convective field coincide with maxima and minima of the stationary probability distribution. For one-dimensional systems, this correspondence is trivially true as $j_{s}(c)$ must vanish on the boundary of and thus everywhere in state space. Bifurcations of the convective field then correspond to qualitative changes in the shape of the stationary probability distribution, so-called phenomenological bifurcations (p-bifurcations)~\cite{Arnold2013}. For two-dimensional systems, however, this correspondence is less clear since recent research has shown that stationary probability currents of the CFPE do not vanish even for reaction networks showing detailed balance~\cite{Ceccato2018}. Still, Mendler et al. suggested that the relation between extrema of the stationary probability distributions and the convective field might also hold in situations where the stationary probability current does not vanish, as they were able to explain the system size-dependent emergence of maxima of stationary probability distributions at the boundary of state space, so-called boundary maxima, for a two-dimensional predator-prey model with help of the convective field. However, this idea has not been systematically explored so far. 

In this paper, we therefore investigate more thoroughly the link between bifurcations of the convective field and p-bifurcations of the corresponding stationary distributions for one- and two-dimensional reaction networks. We focus on saddle-node and Hopf bifurcations. While saddle-node bifurcations occur in one- and two-dimensional systems, Hopf bifurcations cannot occur below two dimensions and always require nonvanishing stationary currents. In the context of Hopf bifurcations, we explore to what extent limit cycles of the convective field correspond to crater-shaped stationary probability distributions. 

Our approach combines two techniques. On the one hand, we derive stability diagrams of the convective field, which we term stochastic stability diagrams. We use them to identify parameter regions for which the convective field makes predictions different from the macroscopic rate equations. In these regions, an agreement between the extrema of the stationary probability distribution and the sources and sinks of the convective field cannot be trivially explained by the macroscopic limit any more. Second, we perform stochastic simulations using Gillespie’s algorithm~\cite{Gillespie1977} to obtain stationary probability distributions. In this way we can check whether topological features of the convective field are correlated with characteristic features of the stationary probability distributions of the reaction networks, both with regard to the shape of the stationary probability distribution as well as qualitative changes of its shape under variation of the system size. Our study uses four different models, which are a one-dimensional positive autoregulator, a two-dimensional double-positive and double-negative feedback loop, and the Brusselator. 


\section{Methods}
\label{sec:Methods}


\subsection{Chemical reaction networks}
\label{ssec:ChemicalReactionNetworks}
A chemical reaction network (CRN) is given a list of chemical reactions for a set of species  X$_{i}$,
\begin{align}
\begin{split}
\ce{$\sigma_{11}$X$_{1}$ + $\sigma_{21}$X$_{2}$ + $...$ + $\sigma_{k1}$X$_{k}$&->[$\mu_{1}$] $\rho_{11}$X$_{1}$ + $\rho_{21}$X$_{2}$ + $...$ $\rho_{k1}$X$_{k}$} \\
&\hspace{11.5pt}\vdots \\
\ce{$\sigma_{1m}$X$_{1}$ + $\sigma_{2m}$X$_{2}$ + $...$ + $\sigma_{km}$X$_{k}$&->[$\mu_{m}$] $\rho_{1m}$X$_{1}$ + $\rho_{2m}$X$_{2}$ + $...$ $\rho_{km}$X$_{k}$} \, .
\label{eq:Reaktionssystem}
\end{split}
\end{align}
The parameters of the reaction network are the stoichiometric constants $\sigma_{ij}$ and $\rho_{ij}$ and the reaction rates $\mu_{i}$. For well-mixed, thermally equilibrated systems the chemical Master equation (CME) corresponding to the reaction system~\eqref{eq:Reaktionssystem} provides a suitable description of its stochastic dynamics~\cite{Gillespie2007, Grima2011, Gillespie1992, Gillespie2009}. However, the CME is analytically intractable for most systems~\cite{Grima2011}. When the number of reactions is not too large, the reaction system \eqref{eq:Reaktionssystem} can be studied by computer simulations using the Gillespie algorithm. For this paper, simulations were performed using the \textit{StochPy} library~\cite{Maarleveld2013} and the software tool \textit{Dizzy}~\cite{Ramsey2005}.

\subsection{Chemical Fokker-Planck equation}
The chemical Fokker-Planck equation (CFPE)~\cite{Gardiner2009} is an often-used approximation to the CME. It is a partial differential equation for the probability density $p(\vec{c}, t)$, 
\begin{equation}
\frac{\partial p(\vec{c},t)}{\partial t} = - \sum_{i} \frac{\partial}{\partial  c_{i}}\left[f_{i}(\vec{c})p(\vec{c},t)\right] + \frac{1}{2\Omega} \sum_{ij} \frac{\partial^{2}}{\partial c_{i}\partial{c_{j}}}\left[D_{ij}(\vec{c})p(\vec{c},t)\right]\, .
\label{eq:CFPE}
\end{equation}
Here,
\begin{equation}
    c_{i} = \frac{n_{i}}{\Omega}
\end{equation}
are the molecular concentrations, with  $\Omega$ denoting the reaction volume and $n_{i}$ the number of molecules of species X$_{i}$. 
The relation between the parameters occuring in the CFPE and those of the original CRN can be expressed in terms of the stoichiometric matrix~\cite{Wilkinson2011} 
\begin{equation}
    S_{ij} = \rho_{ij} - \sigma_{ij}
\label{eq:StoechiometrischeMatrix}
\end{equation}
and the propensity vector~\cite{Schnoerr2017} 
\begin{equation}
    \nu_{j}(\vec{n}, \Omega) = \mu_{j}\prod_{z=1}^{k} \Omega^{-\sigma_{zj}}\cdot\frac{n_{z}!}{(n_{z}-\sigma_{zj})!}.
\label{eq:Propensitaetsvektor}
\end{equation}
The first term on the right-hand side of the CFPE~\eqref{eq:CFPE} contains the drift vector
\begin{equation}
    \vec{f}(\vec{c}) = \textbf{\textit{S}}\cdot\vec{\nu}(\vec{c})\, .
\end{equation}
It results from the deterministic part of the reaction system. This deterministic part gives the macroscopic rate equations for the reaction network~\eqref{eq:Reaktionssystem},
\begin{equation}
\frac{\text{d}\vec{c}}{\text{d}t} = \vec{f}(\vec{c}) .
\label{eq:DeterministischeReaktionskinetik}
\end{equation}
The second term on the right-hand side of the CFPE~\eqref{eq:CFPE} contains the diffusion matrix
\begin{equation}
    \textbf{\textit{D}}(\vec{c}) = \textbf{\textit{S}}\cdot\text{diag}(\vec{\nu})\cdot\textbf{\textit{S}}^{T},
\end{equation}
which is due to the stochastic fluctuations of the concentrations. 

\subsection{The convective field and the stationary probability distribution}
The CFPE has the form of a continuity equation
\begin{equation}
\frac{\partial p(\vec{c},t)}{\partial t} = - \vec{\nabla}\cdot\vec{j}(\vec{c},t)\, .
\label{eq:Kontinuitaetsgleichung}
\end{equation}
Defining 
\begin{equation}
\vec{\alpha}(\vec{c}) = \vec{f}(\vec{c}) - \frac{1}{2\Omega} \sum_{ik}\frac{\partial D_{ik}}{\partial c_{k}}\vec{\epsilon}_{i}\, ,
\label{eq:Konvektionsfeld}
\end{equation}
with $\vec{\epsilon}_{i}$ being the unit vector in direction $c_{i}$, the probability current can be written as 
\begin{equation}
\vec{j}(\vec{c},t) = \underbrace{\vec{\alpha}(\vec{c})p(\vec{c},t)}_{\vec{j}_{c}(\vec{c},t)} - \frac{1}{2\Omega}\underbrace{\textbf{\textit{D}}(\vec{c})\cdot\vec{\nabla}p(\vec{c},t)}_{\vec{j}_{d}(\vec{c},t)}
\label{eq:WahrscheinlichkeitsstromdichteFull}
\end{equation}
with a convective current $\vec{j}_{c}(\vec{c},t)$ and a diffusive current $\vec{j}_{d}(\vec{c},t)$~\cite{Mendler2018}. The convective current describes a directed motion through state space that is not caused by concentration gradients. Apart from the deterministic drift term, it contains also a contribution that is due to concentration dependence of the diffusion matrix.
For large times, the probability distribution $p(\vec{c},t)$ approaches a stationary distribution $p_{s}(\vec{c})$~\cite{vanKampen1992}. Since the stationary distribution does not change in time, it follows from~\eqref{eq:Kontinuitaetsgleichung} that the stationary probability current  $\vec j_s(\vec c)$ satisfies 
\begin{equation}
\vec{\nabla}\cdot\vec j_s(\vec{c}) = 0 \, .
\label{eq:stationarycurrent}
\end{equation}
For one-dimensional systems, the general solution of this condition is a constant stationary probability current. With closed boundary conditions, there can be no current through the boundary. The stationary current must therefore vanish everywhere in one-dimensional systems, and from~\eqref{eq:WahrscheinlichkeitsstromdichteFull} follows then that the convective field $\vec{\alpha}(\vec{c})$ vanishes at maxima and minima of $p_{s}(\vec{c})$~ \cite{Mendler2018}. Mendler et al.~define favorable states of a stochastic system as maxima of the stationary probability distribution $p_{s}(\vec{c})$ and unfavorable states as minima of $p_s{}(\vec{c})$. With this definition, favorable and unfavorable states correspond to sinks and sources, i.e. fixed points, of $\vec{\alpha}(\vec{c})$.

For higher-dimensional systems, the correspondence between extrema of stationary probability distributions and sources and sinks of $\vec{\alpha}(\vec{c})$ holds strictly only under the condition that  $\vec j_s(\vec{c})$ vanishes. The (un)favorable states can then be found from the fixed-point condition of the convective field
\begin{equation}
\vec{\alpha}(\vec{c}) = \vec{f}(\vec{c}) - \frac{1}{2\Omega} \sum_{ik}\frac{\partial D_{ik}}{\partial c_{k}}\vec{\epsilon}_{i} = 0.
\label{eq:FPCondAlpha}
\end{equation}
In higher-dimensional systems, stationary probability currents do not vanish in general~\cite{Ceccato2018}. In this case, the correspondence between extrema of the stationary probability distribution and the fixed points of the convective field can only be approximately valid. This approximation, however, becomes exact in the limit of infinite system size, where extrema of $p_s(\vec{c})$ must coincide with fixed points of $\vec f(\vec{c})$, and $\vec{f}(\vec{c})$ in turn coincides with $\vec\alpha(\vec{c})$. 

\subsection{Bifurcations}
When the number of fixed points or their stability changes, a dynamical system undergoes a bifurcation. Due to the additional term that depends on the diffusion matrix, the bifurcations of the convective field $\vec \alpha(\vec c)$ are shifted in parameter space relative to those of the macroscopic system described by $\vec f(\vec c$).

Whenever fixed points of the convective field correspond to nearby extrema of the stationary probability distribution, the bifurcations of the convective field are accompanied by according changes in the extrema of the stationary probability distribution. Such qualitative changes in the structure of the maxima and minima of the stationary probability distribution are  so-called phenomenological bifurcations or p-bifurcations~\cite{Arnold2013}. 

In this paper, we focus on two types of bifurcations that occur in two-dimensional systems: In a saddle-node bifurcation, a stable and an unstable fixed point are destroyed or created as a control parameter changes. The corresponding phenomenological bifurcation is the merging or creation of a local probability maximum and minimum (1D) or saddle (2D). In a Hopf bifurcation, a stable fixed point becomes unstable, and a limit cycle is created. In the corresponding phenomenological bifurcation a local probability maximum turns into a crater, with a local probability minimum that is surrounded by a ridge. 

An important tool for our investigation will be stochastic stability diagrams, i.e. stability diagrams of the convective field. They give a concise qualitative overview of the behavior of the convective field in dependence of control parameters. We will compare the behavior of the fixed points of the convective field to that of the stationary probability distribution, which we will obtain by stochastic simulations of the reaction system \eqref{eq:Reaktionssystem}. 

The most interesting regions in parameter space are those where the stochastic stability diagram deviates from that of the macroscopic model. Such a deviation means that the convective field undergoes a bifurcation when the system size is changed and all other parameters remain fixed. We will focus on these parameter regions in order to explore to what extent the sources and sinks of the convective field correlate with extrema of the stationary probability distribution.


\section{Results}
\label{sec:Results}


\subsection{A bistable one-dimensional system: Positive autoregulator}
\label{ssec:PAR}
One of the simplest reaction systems showing bistable behavior is a positive autoregulator~\cite{Alon2007, Shoval2010}. An autoregulator consists of a single gene encoding a transcriptional factor (protein) that acts as an activator of that gene. Since the dynamics of the mRNA concentration is for many organisms much faster than that of the protein concentration, we assume that it is in equilibrium with the protein concentration. Then the dynamics of the autoactivator can be described by the following three reactions for the protein X: 
\begin{align}
\begin{split}
\ce{\emptyset &->[$b$] X} \, ,\qquad \ce{$\emptyset$ ->[$\frac{mx^{n}}{\theta^{n} + x^{n}}$] X} \, , \qquad \ce{X ->[$1$] \emptyset}\, .
\end{split}
\label{eq:ReaktionsgleichungenPAR}
\end{align}
These reactions describe basal production with constant rate $b$, autoactivation with a concentration-dependent rate, and  loss of X through dilution or active degradation with unit rate. Autoactivation is implemented by using a Hill function~\cite{Hill1913} with maximum production rate $m$, half-saturation constant $\theta$, and the Hill coefficient $n$ as parameters.

We first analyze the macroscopic rate equations for the concentration $x$ of protein X. For the autoactivator, equation \eqref{eq:DeterministischeReaktionskinetik} takes the form
\begin{equation}
\dot{x} = b + \frac{mx^{n}}{\theta^{n}+x^{n}} - x\, .
\label{eq:DeterministischeKinetikPAR}
\end{equation}
Using the dimensionless variables
\begin{equation}
    \xi = \frac{x}{\theta}\, ,\qquad \beta = \frac{b}{\theta}\, ,\qquad \mu = \frac{m}{\theta}\, ,
\end{equation}
equation~\eqref{eq:DeterministischeKinetikPAR} becomes
\begin{equation}
\dot{\xi} = \beta + \frac{\mu\xi^{n}}{1+\xi^{n}} - \xi\, .
\label{eq:DimensionsloseKinetikPAR}
\end{equation}
The parameters $\beta$ and $\mu$ quantify the importance of basal production and production through feedback relative to the influence of degradation. 

\begin{figure}[!htbp]
\centering
\subfigure{\includegraphics[height=0.25\textheight]{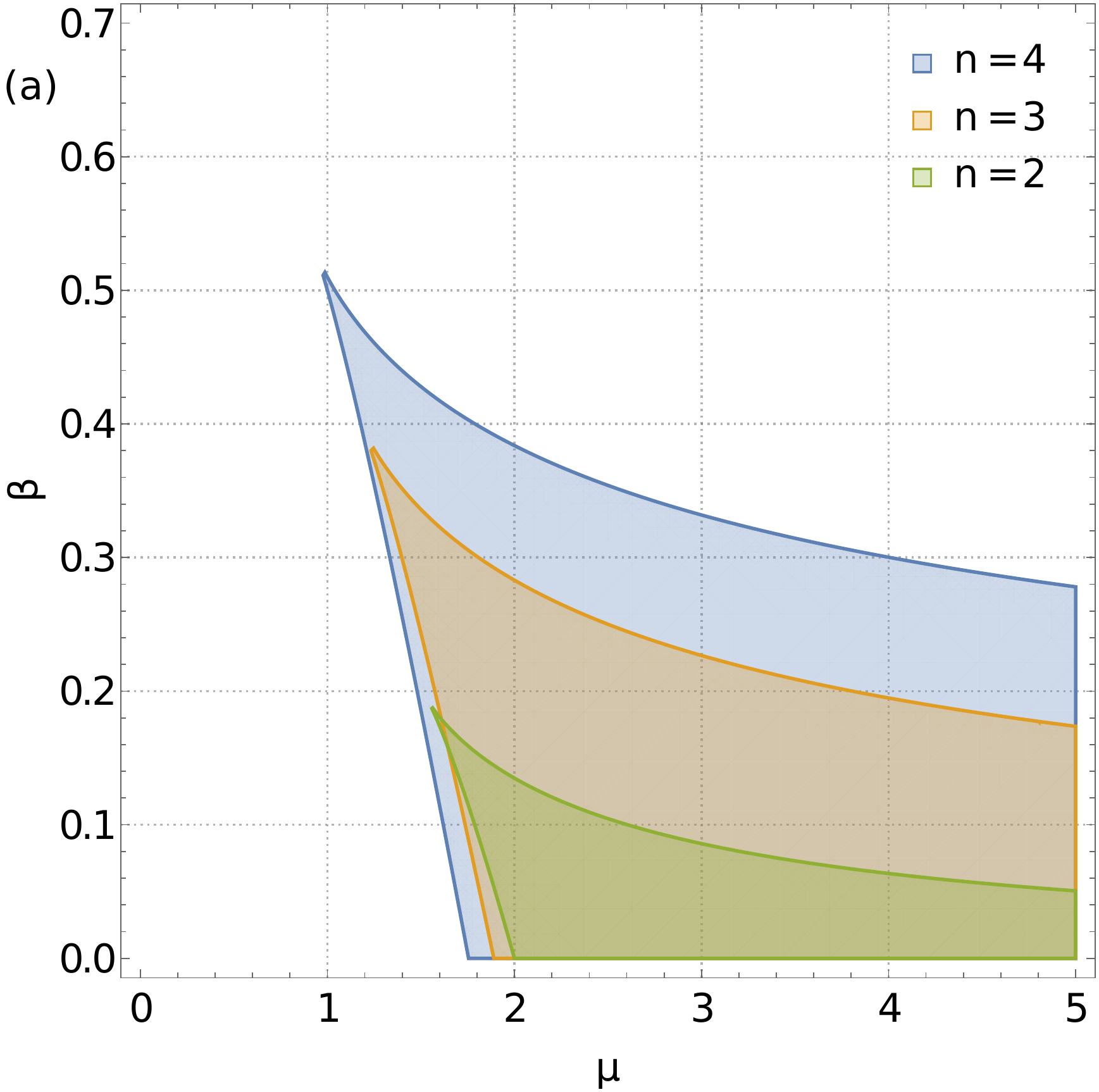}} \qquad \qquad
\subfigure{\includegraphics[height=0.25\textheight]{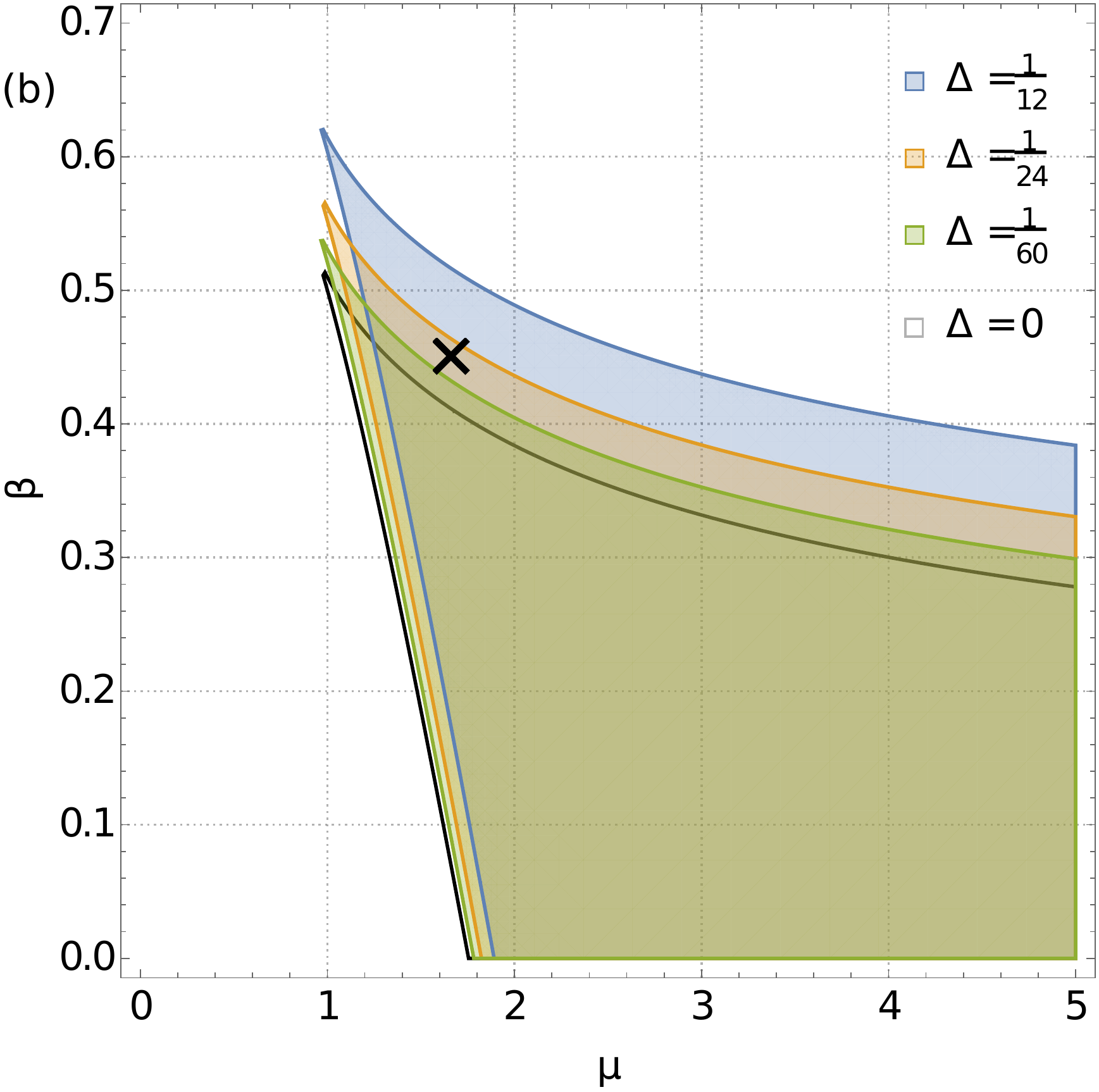}}
\caption{(a) Stability diagram of the macroscopic autoactivator model~\eqref{eq:DimensionsloseKinetikPAR} for different values of the Hill coefficient $n$. Within the shaded region, the system has two stable steady states. This bistable region becomes larger with increasing $n$. Outside of it, the system is monostable. The boundaries mark saddle-node bifurcations. (b) Stochastic stability diagram of the autoactivator model \eqref{eq:ReduziertesKonvektionsfeldPAR} for $n = 4$, $\theta = 6$ and various values of the discreteness parameter $\Delta = \frac{1}{2\Omega\theta}$. With increasing system size $\Omega$, the stability diagram of the macroscopic system is approached. The position of the system depicted in figure~\ref{fig:figure2} in parameter space is marked with a cross.} 
\label{fig:figure1}
\end{figure}

A saddle-node bifurcation of~\eqref{eq:DimensionsloseKinetikPAR} occurs for parameter values such that the functions $(\xi - \beta)$ and $\mu\xi^{n}/(1+\xi^{n})$ are tangent to each other. Mathematically, this translates to the condition $(\partial_{\xi}\dot{\xi})\vert_{\xi=\xi^{*}} = 0$, together with the fixed point condition $\dot\xi=0$ for $\xi^{*}$. These conditions can be used to parametrize the bifurcation lines in parameter space. The resulting stability diagram is shown in figure~\ref{fig:figure1}(a). 
Now let us turn to the stochastic version of model~\eqref{eq:ReaktionsgleichungenPAR} and investigate the convective field \eqref{eq:Konvektionsfeld},
\begin{equation}
\alpha(x) = b + \frac{mx^{n}}{\theta^{n} + x^{n}} - x - \frac{1}{2\Omega}\left(\frac{\text{d}}{\text{d}x}\frac{mx^{n}}{\theta^{n} + x^{n}} + 1 \right).
\label{eq:KonvektionsfeldPAR}
\end{equation}
By introducing the dimensionless convective field $\tilde \alpha=\alpha/\theta$ and the discreteness parameter~\cite{Scott2007} $\Delta = 1/2\Omega\theta$, we obtain
\begin{equation}
\tilde{\alpha}(\xi) = \beta + \frac{\mu\xi^{n}}{1+\xi^{n}} - \xi - \Delta\left(\frac{n\xi^{n-1}}{(1+\xi^{n})^{2}} + 1\right).
\label{eq:ReduziertesKonvektionsfeldPAR}
\end{equation}
The discreteness parameter $\Delta$ scales inversely with the number of molecules $N_{A} = \Omega\theta$ needed to activate production of X through autoregulation. As $N_{A}$ becomes smaller, the last term of \eqref{eq:ReduziertesKonvektionsfeldPAR}, which is due to intrinsic fluctuations and describes the deviation from the macroscopic model, becomes more important. 

Again, we obtain the bifurcation lines of the dynamical system~\eqref{eq:ReduziertesKonvektionsfeldPAR} by requiring that $(\partial_{\xi}\tilde{\alpha})\vert_{\xi=\xi_{\alpha}^{*}} = 0$, where $\xi_{\alpha}^{*}$ is a solution of $\tilde{\alpha} = 0$. This gives the stochastic stability diagram shown in figure~\ref{fig:figure1}(b) for $n=4$. The larger $\Delta$, i.e., the smaller $\Omega$, the larger is the deviation from the stability diagram of the macroscopic model. 

\begin{figure}[!htbp]
\centering
\subfigure{\includegraphics[width=1.0\columnwidth]{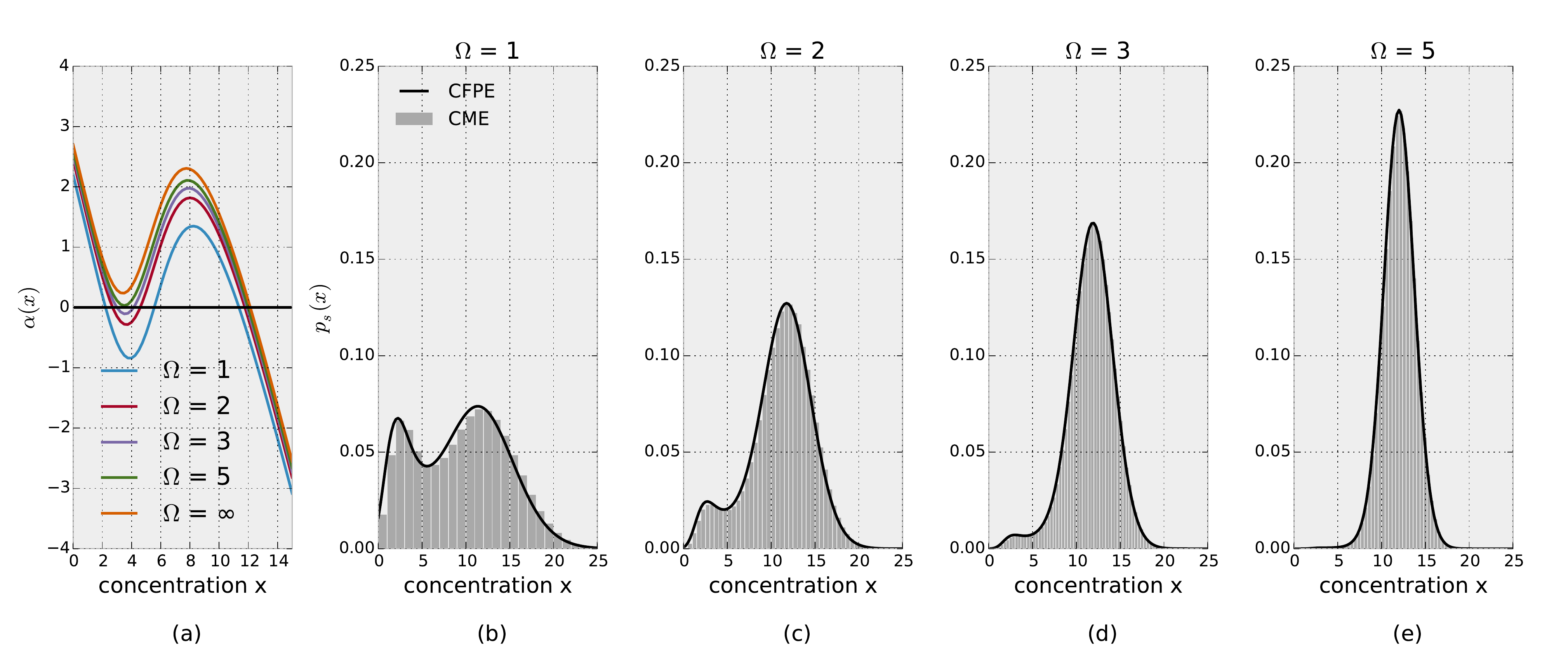}}
\hspace{-0.3 cm}
\caption{(a) Plot of $\alpha(x)$ for the autoactivator \eqref{eq:KonvektionsfeldPAR} for various system sizes $\Omega$ with parameters $n = 4$, $\theta = 6$, $m = 10$, $b = 2.7$. The zeroes of $\alpha$ correspond to the fixed points of the convective field. Their number makes a transition from three to one as the system size $\Omega$ is increased. (b) - (e) Simulated stationary probability distributions corresponding to the four finite system sizes shown in figure~\ref{fig:figure2}(a). A phenomenological bifurcation from bimodal to unimodal behavior occurs with increasing system size. For comparison, the stationary solutions of the corresponding CFPEs are shown as solid lines.}
\label{fig:figure2}
\end{figure}

Figure~\ref{fig:figure2} demonstrates that the fixed points of $\alpha(x)$ agree with the extrema of the stationary probability distribution for the system marked in figure~\ref{fig:figure1}(b) by a cross. In contrast to the previous figure, we now use the system size $\Omega$ as the bifurcation parameter. The figure shows that the saddle-node bifurcation of the convective field corresponds to a transition from a bimodal to a unimodal stationary probability distribution. From figure~\ref{fig:figure1}(b) it is evident that the opposite case, a transition from a unimodal to a bimodal stationary distribution with increasing $\Omega$, occurs also in this model, but it is not shown here.  

The system-size dependence of these bifurcations follows from the properties of the diffusion coefficient
\begin{equation}
    D(x) = b + \frac{mx^{n}}{\theta^{n} + x^{n}} + x\, ,
\end{equation}
the derivative of which, divided by twice the system size, determines the difference between $f(x)$ and $\alpha(x)$, cf.~equation~\eqref{eq:FPCondAlpha}. The diffusion coefficient $D(x)$ increases with $x$, which means that $\alpha(x) < f(x)$ everywhere, as can be seen in figure \ref{fig:figure2}(a). Consequently, the zeroes of $\alpha(x)$ are left of those of $f(x)$ when the zero is associated with a stable fixed point (negative slope of $f$), and right of those of $f(x)$ when the zero is associated with an unstable fixed point. This means that maxima of the stationary probability distribution are shifted to the left and minima to the right when the system size becomes smaller. As a consequence, we see in figure \ref{fig:figure2}(b) that with increasing system size the left maximum moves right and collides with the minimum, which moves left. 

\begin{figure}[!htbp]
\centering
\subfigure{\includegraphics[width=1.0\columnwidth]{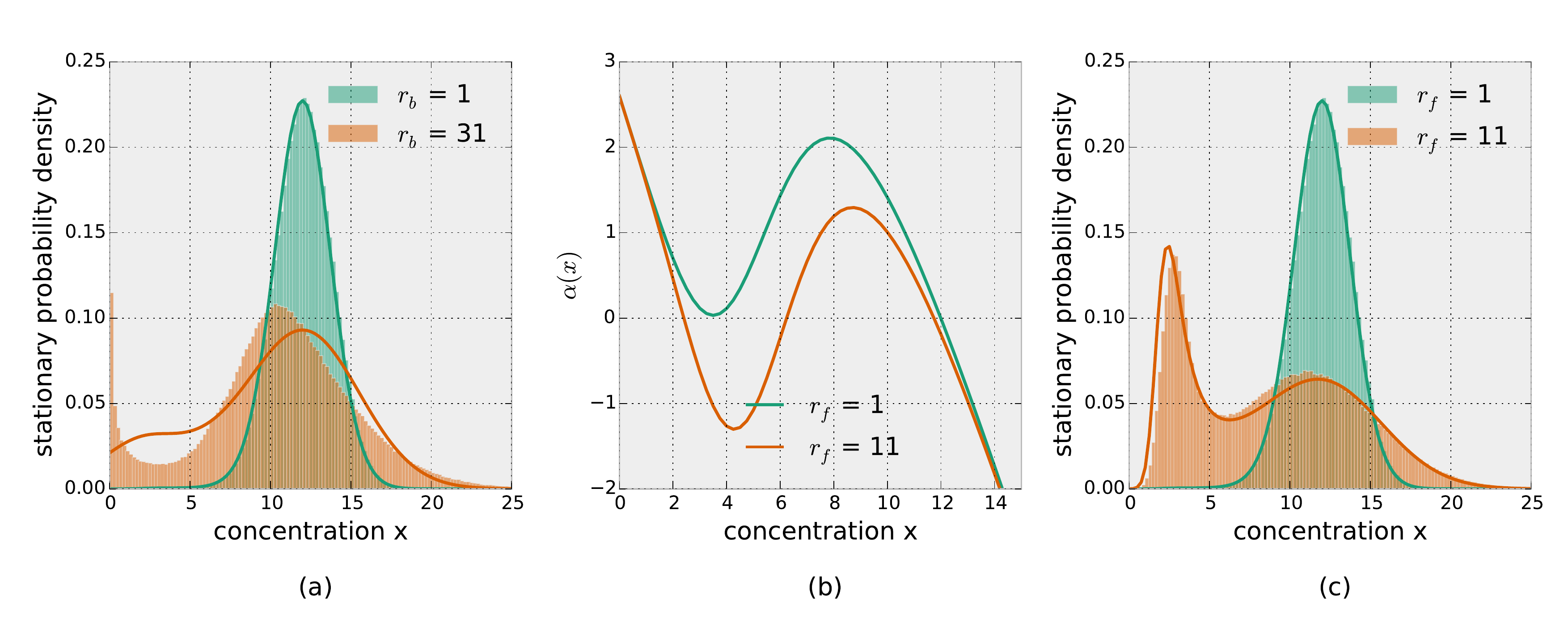}}
\caption{(a) Simulated stationary probability distributions for the system shown in figure~\ref{fig:figure2}(e) and two different basal burst parameters $r_{b}$. A boundary maximum emerges for large values of $r_{b}$ as introduced in equation~\eqref{eq:BurstNoise_PAR}. Stationary solutions of the CFPE are shown as solid lines for comparison. (b) Plots of the convective field of the autoactivator for the system shown in figure~\ref{fig:figure2}(e) and for two values of the burst parameter $r_{f}$. Compared to figure~\ref{fig:figure3}(a), implementation of reaction~\eqref{eq:BurstNoise_PAR_2} into the original model now leads to a change in the number of zeroes of the convective field with $r_{f}$. (c) Simulated stationary probability distributions for the autoactivator with bursty production through feedback. The parameters are chosen as in figure~\ref{fig:figure2}(e). Maxima of the stationary probability distributions can emerge or vanish under variation of $r_{f}$. Stationary solutions of the CFPE are shown as solid lines for comparison.}
\label{fig:figure3}
\end{figure}

While figure \ref{fig:figure2} shows that the agreement between the stationary solution of the CME and of the CFPE is very good even for small system sizes, there exist variants of the model where this agreement breaks down. In this case, the phenomenological bifurcations of the stationary probability distribution of the CME do not agree any more with those of the convective field $\alpha(x)$. One example of this is burst noise, which is a common form of stochasticity in gene transcription and translation~\cite{Bokes2017, Raj2008}. Here, molecule numbers change by at least two in a single reaction event. When we implement burst noise for the autoactivator by introducing a burst parameter $r_{b}$ and replacing basal production in~\eqref{eq:ReaktionsgleichungenPAR} by 
\begin{align}
\begin{split}
\ce{\emptyset &->[$b/r_{b}$] r$_{\text{b}}$X}\, ,
\end{split}
\label{eq:BurstNoise_PAR}
\end{align}
the convective field is left unchanged. The stationary probability distribution of the CFPE merely broadens, as this type of noise makes an $x$-independent contribution to the diffusion coefficient of the CFPE. The CME, however, can develop a bimodal stationary probability distribution when $r_{b}$ is chosen sufficiently large. Both effects are shown in figure~\ref{fig:figure3}(a). 

In contrast, when we implement bursty production through feedback by setting
\begin{align}
\begin{split}
\ce{\emptyset &->[$\frac{1}{r_{f}}\cdot\frac{mx^{n}}{\theta^{n} + x^{n}}$] r$_{\text{f}}$X}
\end{split}
\label{eq:BurstNoise_PAR_2}
\end{align}
in the original model~\eqref{eq:ReaktionsgleichungenPAR}, burst noise of the autoactivator is reflected in the convective field, see figure~\ref{fig:figure3}(b). The contribution of $r_{f}$ to the convective field does not vanish in this case, as can be seen from~\eqref{eq:Konvektionsfeld}. In this case, we find a correspondence between saddle-node bifurcations of the convective field and p-saddle-node bifurcations due to burst noise, see figure~\ref{fig:figure3}(c). 


\subsection{A bistable two-dimensional system: Positive feedback loop}
\label{ssec:PFL}

Next, we study a two-dimensional reaction system capable of bistable behavior: the positive feedback loop. Often found in developmental transcription networks, positive feedback loops consist of two molecular species either both activating or both repressing each other~\cite{Shoval2010}. They are called double-positive and double-negative feedback loops, respectively~\cite{Alon2007}. We will examine both versions of the positive feedback loop in the following. 

\subsubsection{Double-negative loop}
\label{sssec:DoubleNegative}
A simple model of a double-negative loop is the reaction system
\begin{align}
\begin{split}
\ce{$\emptyset$ ->[$\frac{m_{x}\theta_{y}^{n}}{\theta_{y}^{n} + y^{n}}$] X} \qquad \ce{X ->[$1$] \emptyset} \\
\ce{$\emptyset$ ->[$\frac{m_{y}\theta_{x}^{n}}{\theta_{x}^{n} + x^{n}}$] Y} \qquad \ce{Y ->[$d$] \emptyset}
\end{split}
\label{eq:ReaktionsgleichungenPFLnn}
\end{align}
of two protein species X and Y. We allow for different degradation rates $d_x=1$ and $d_y=d$, different maximum transcription rates $m_{i}$ and different activation thresholds $\theta_{i}$, but assume for simplicity that the Hill coefficient $n$ is identical for both species.  Since the two proteins repress each other, this model shows bistable behavior with one of the proteins having a high and the other a low concentration.

\begin{figure}[!htbp]
\centering
\subfigure{\includegraphics[height=0.25\textheight]{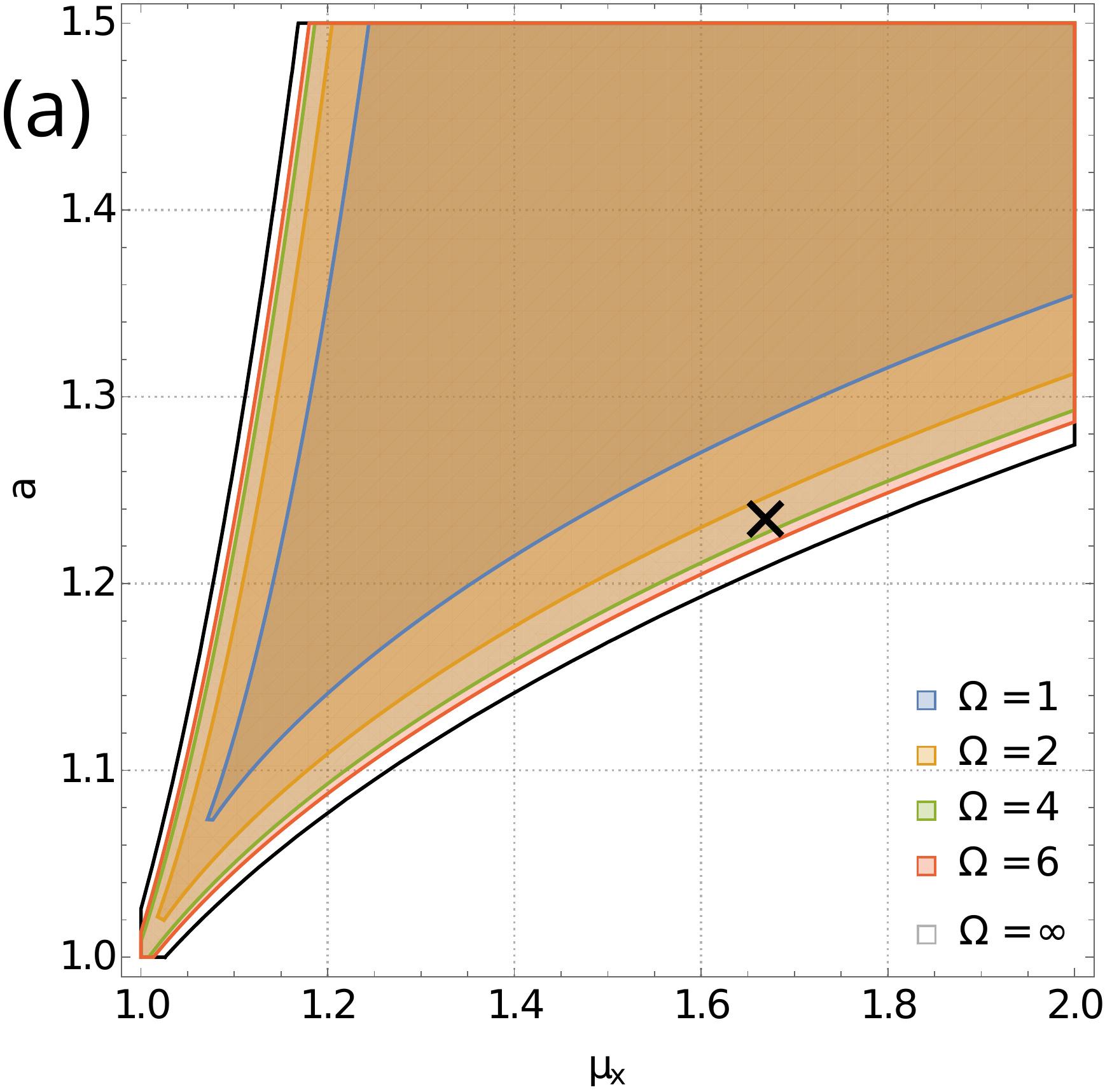}} \qquad \quad
\subfigure{\includegraphics[height=0.25\textheight]{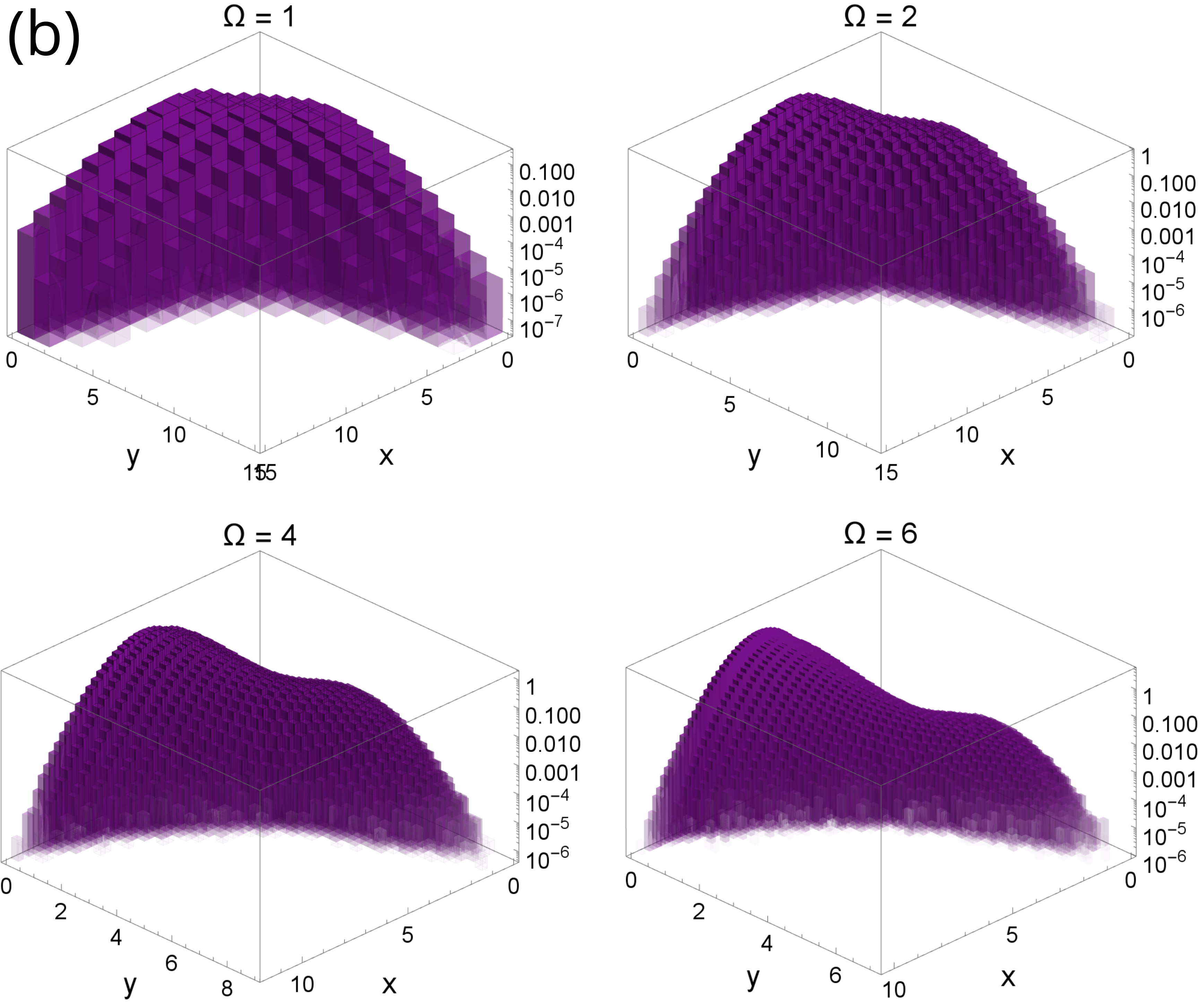}}
\caption{(a) Stability diagram of the double-negative loop for  $n = 5$, $\theta_{x} = 3$, $\theta_{y} = 3$,  $d = 2.35$ and various system sizes $\Omega$. The parameter values $m_{x} = 5$ and $m_{y} = 8.7$ for the stochastic simulations are marked with a black cross. For system sizes below $\Omega_{c} \approx 2.94$ the system falls into the monostable region of the convective field. (b) Simulated stationary probability distributions of the double-negative feedback loop for the parameter values indicated by the cross on the left, and for four different values of $\Omega$. For larger system sizes, the stationary probability distributions attain a bimodal shape. This correlates with a saddle-node bifurcation of the convective field. Note that the histograms are plotted logarithmically, as the second mode would be hard to detect on a linear scale. 
} 
\label{fig:figure4}
\end{figure}

\begin{figure}[!htbp]
\centering
\subfigure{\includegraphics[height=0.25\textheight]{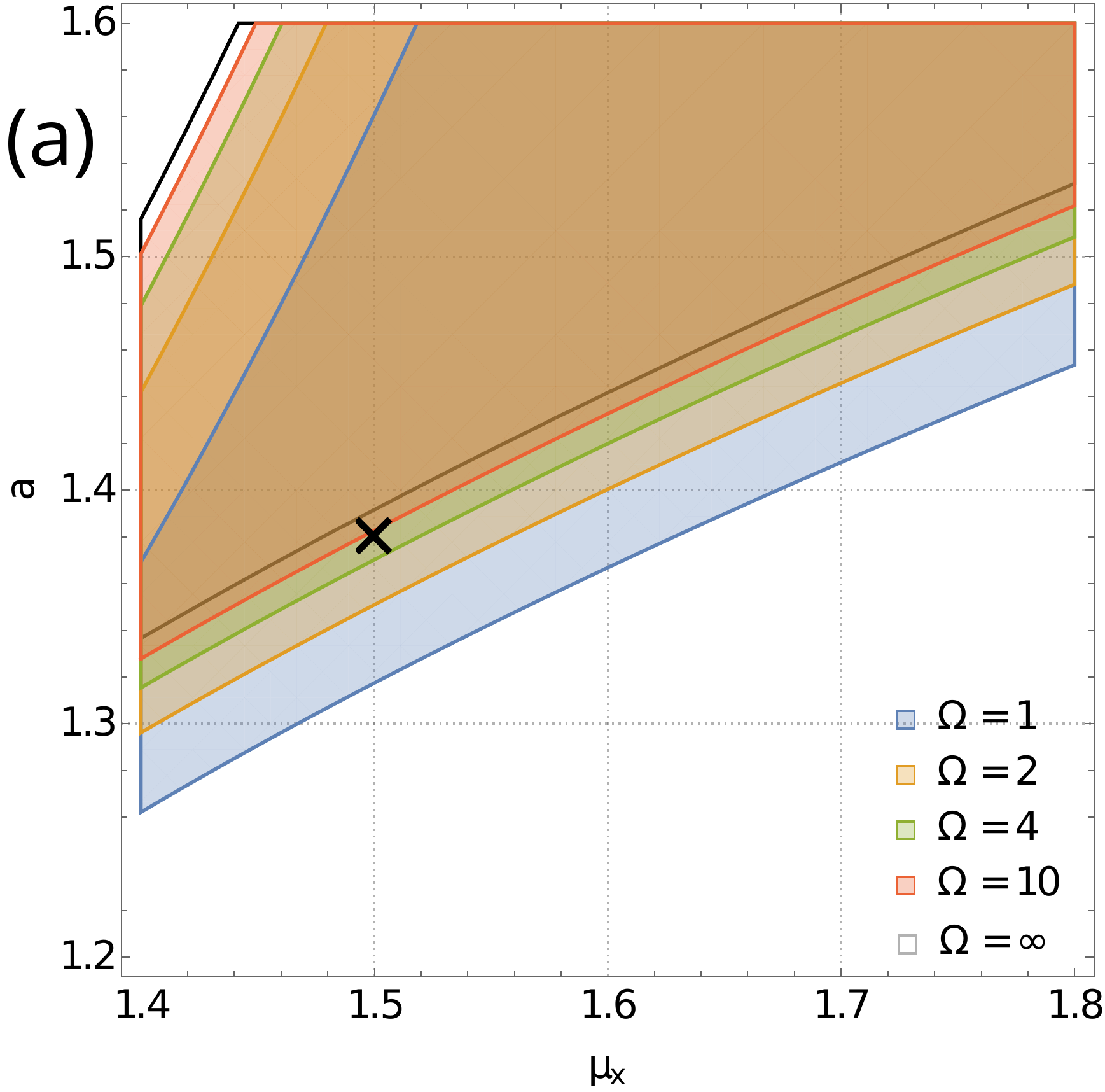}} \qquad \quad
\subfigure{\includegraphics[height=0.25\textheight]{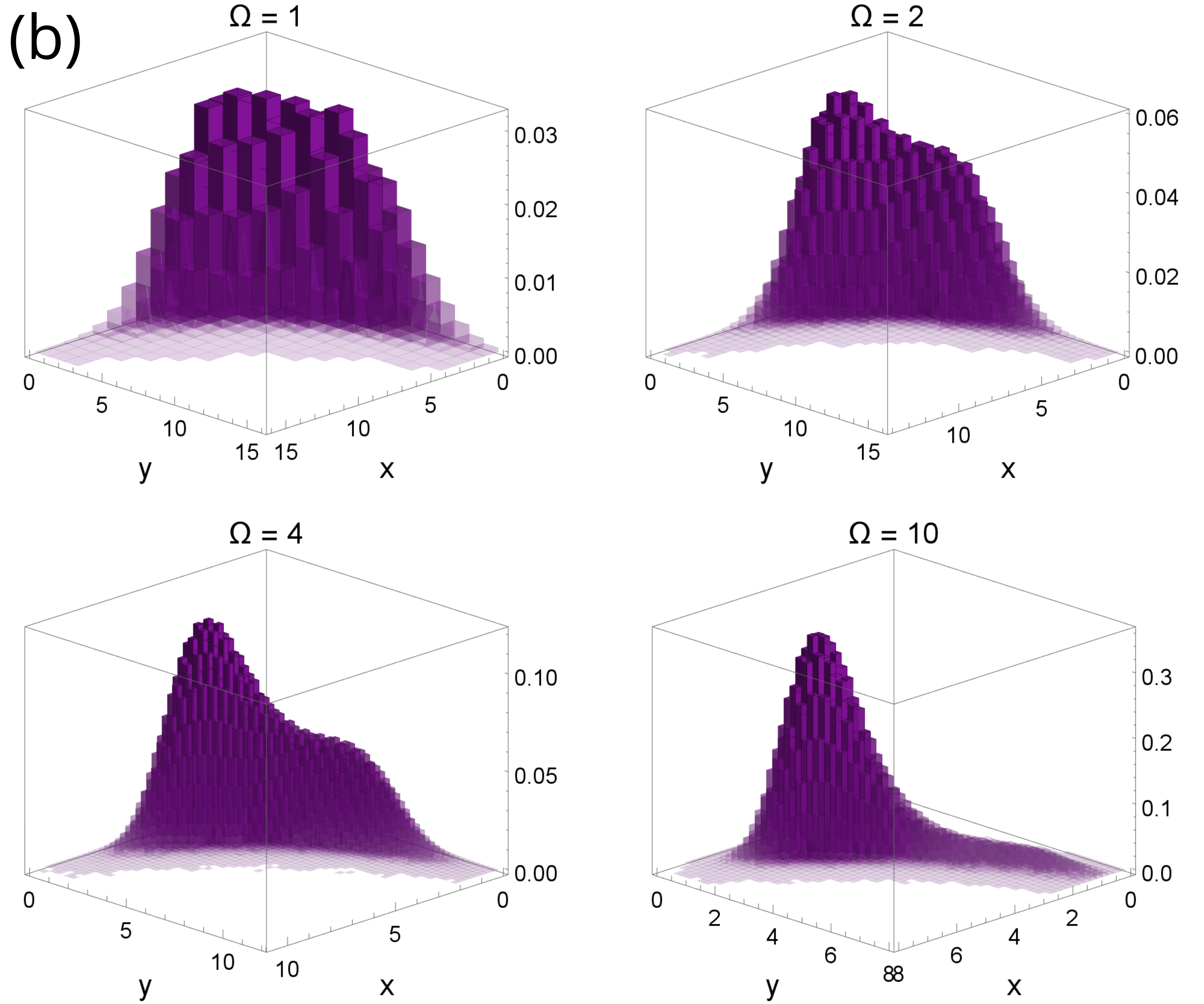}}
\caption{(a) Stability diagram of the double-negative loop for $n = 3$, $\theta_{x} = 3$, $\theta_{y} = 5$, $d = 1$ and various system sizes $\Omega$. The parameter values $m_{x} = 4.5$ and $m_{y} = 6.9$ for the stochastic simulations are marked with a black cross. For $\Omega < \Omega_{c} \approx 7.55$ the system falls into the bistable region of the convective field. Above $\Omega_{c}$, the convective field exhibits a single stable fixed point, just as the macroscopic system. (b) Simulated stationary probability distributions of the double-negative feedback loop for the parameter values indicated by the cross on the left, and for four different values of $\Omega$. The histograms show only one maximum except for $\Omega = 1$, while the convective field shows bistability for the three smallest system sizes.} 
\label{fig:figure5}
\end{figure}

The convective field of the reaction system~\eqref{eq:ReaktionsgleichungenPFLnn} reads
\begin{equation}
\renewcommand\arraystretch{1.4}
\vec{\alpha}(x,y) = \begin{pmatrix}m_{x}\frac{\theta_{y}^{n}}{\theta_{y}^{n} + y^{n}} - x\\ m_{y}\frac{\theta_{x}^{n}}{\theta_{x}^{n} + x^{n}} -dy\end{pmatrix} - \frac{1}{2\Omega}\begin{pmatrix} 1 \\ d \end{pmatrix}\, .
\label{eq:KonvektionsfeldPFLnn}
\end{equation}
We switch again to dimensionless variables 
\begin{equation}
    \tilde{\alpha}_{i} = \frac{\alpha_{i}}{\theta_{i}}\, , \qquad \xi = \frac{x}{\theta_{x}}\, , \qquad \upsilon = \frac{y}{\theta_{y}}\, , \qquad \mu_{i} = \frac{m_{i}}{\theta_{i}}\, , \qquad \Delta_{i} = \frac{1}{2\Omega\theta_{i}}
\label{eq:VariablenwechselPFLnn}
\end{equation}
and the dimensionless convective field  
\begin{equation}
\renewcommand\arraystretch{1.1}
\vec{\tilde{\alpha}}(\xi,\upsilon) = \begin{pmatrix}\mu_{x}\frac{1}{1 + \upsilon^{n}} - \xi\\ \mu_{y}\frac{1}{1 + \xi^{n}} -d\upsilon\end{pmatrix} - \begin{pmatrix} \Delta_{x} \\ d\Delta_{y} \end{pmatrix}\text{.}
\label{eq:KonvektionsfeldPFLnn_rescaled}
\end{equation}
In order to obtain the stability diagram, we derive from the fixed point condition 
$\vec{\tilde{\alpha}}(\xi^{*}, \upsilon^{*}) = 0$
a self-consistency equation 
\begin{equation}
    \upsilon^{*} = \frac{a}{1 + \xi^{*}(\upsilon^{*})^{n}} - \Delta_{y}
    \label{eq:SelbstkonsistenzgleichungPFLnn}
\end{equation}
for the fixed point value of $\upsilon$, where $a$ is defined as
\begin{equation}
    a = \frac{\mu_{y}}{d}\, .
\label{eq:PFLnnDefinition_a}
\end{equation}
The solution set of equation~\eqref{eq:SelbstkonsistenzgleichungPFLnn} depends on the five parameters $\mu_{x}$, $a$, $n$, $\Delta_{x}$, and $\Delta_{y}$.
We determined numerically the regions in parameter space where the relation~\eqref{eq:SelbstkonsistenzgleichungPFLnn} has 1 or 3 solutions. Examples of stability diagrams obtained in this way are shown in figures~\ref{fig:figure4}(a) and \ref{fig:figure5}(a). 

In figure \ref{fig:figure4}(a), the two discreteness parameters are identical, and the bistable region is shifted along the identity line with changing system size. Therefore, only a transition from monostable to bistable behavior is observed with increasing system size. In figure \ref{fig:figure5}(a), the two discreteness parameters are different, and the opposite transition from bistable to monostable behavior with increasing system size occurs also. 

Figures \ref{fig:figure4}(b) and \ref{fig:figure5}(b) show how the stationary probability distributions of the reaction network~\eqref{eq:ReaktionsgleichungenPFLnn} change as the system size moves through the transition between monostable and bistable behavior of the convective field. In figure \ref{fig:figure4}(b), we find good agreement between the bifurcation of the convective field and the number of peaks of the stationary probability distribution: For system sizes below $\Omega_{c} \approx 2.94$, where the convective field has only one stable fixed point, the stationary probability distribution does not show two distinct peaks, but it does for larger system sizes. The bifurcation point of the simulated system though is difficult to identify. This is because the relative weight of the two peaks depends on $\Omega$ and can vary greatly, making bimodality hard to detect.

In figure \ref{fig:figure5}(b), only the probability distribution for $\Omega = 1$ exhibits a bimodal shape, although three of the simulated stationary probability distributions were obtained for parameters where the convection field shows bistability. But even for $\Omega=1$ the two peaks cannot be clearly distinguished. We thus see that bistability of the convective field does not necessarily imply the existence of two peaks of the stationary distribution of simulated systems, and that the bifurcation point $\Omega_{c}$ of the convective field need not coincide a the p-bifurcation of the corresponding reaction network. Still, the stationary distributions shown in figure~\ref{fig:figure5}(b) show a shoulder where the second stable fixed point of $\vec{\alpha}(x,y)$ is located. This shoulder shrinks as system size grows, as would happen for a stochastic system moving away from a saddle-node bifurcation. 

\subsubsection{Double-positive loop}
\label{sssec:DoublePositive}
The chemical reactions of the double-positive loop are
\begin{align}
\begin{split}
&\ce{$\emptyset$ ->[$b_{x}$] X} \qquad \ce{$\emptyset$ ->[$\frac{m_{x}\theta_{y}^{n_{y}}}{\theta_{y}^{n_{y}} + y^{n_{y}}}$] X} \qquad \ce{X ->[$1$] \emptyset} \\
&\ce{$\emptyset$ ->[$b_{y}$] Y} \qquad \ce{$\emptyset$ ->[$\frac{m_{y}\theta_{x}^{n_{x}}}{\theta_{x}^{n_{x}} + x^{n_{x}}}$] Y} \qquad \ce{Y ->[$d$] \emptyset}\text{ .}
\end{split}
\label{eq:ReaktionsgleichungenPFLpp}
\end{align}
In contrast to the previous model, we included a basal expression of X and Y with rates $b_{x}$ and $b_{y}$. This is necessary because otherwise the complete lack of X and Y would be an absorbing state. We now allow for different values for the Hill coefficients $n_{x}$ and $n_{y}$ to obtain a larger extent of asymmetry in the system. A completely symmetric system would be effectively one-dimensional since the dynamics of the dynamical system described by $\vec\alpha(x,y)$ would then be attracted to the diagonal $x=y$, on which all fixed points are located, and along which the saddle-node bifurcations occur.%

\begin{figure}[!htbp]
\centering
\subfigure{\includegraphics[height=0.25\textheight]{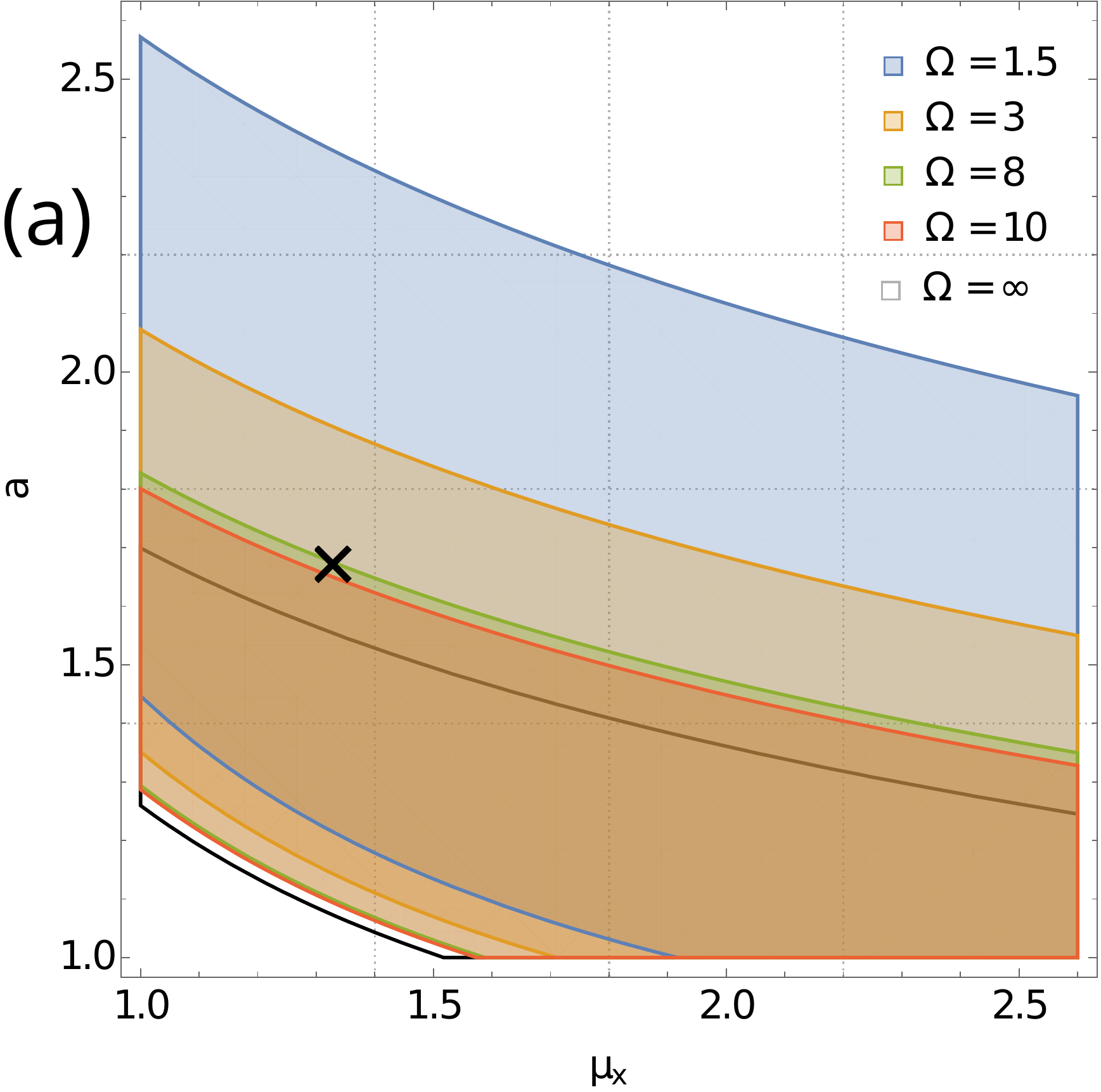}} \qquad \quad
\subfigure{\includegraphics[height=0.25\textheight]{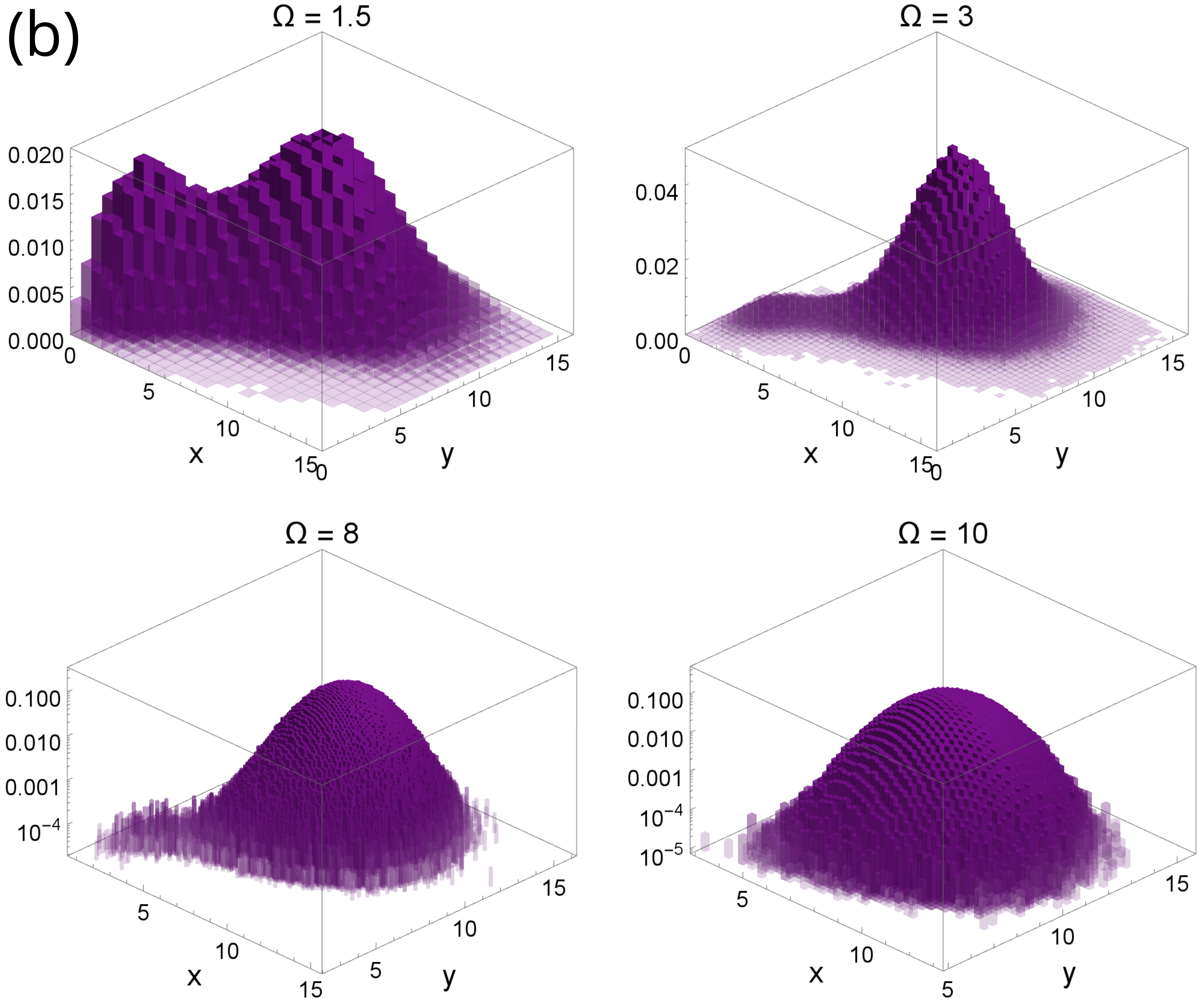}}
\caption{(a) Stability diagram of the double-positive loop for $n_{x} = 2$, $n_{y} = 8$, $\theta_{x} = 5$, $\theta_{y} = 6$,  $d = \frac{2}{3}$, $b_{x} = 2$, $b_{y} = \frac{5}{3}$, and various system sizes $\Omega$. The parameter values $m_{x} = m_{y} = \frac{20}{3}$ for the stochastic simulations are marked with a black cross. For $\Omega < \Omega_{c} \approx 8.28$ the convective field shows bistability for these parameter values. 
(b) Simulated stationary probability distributions of the double-positive feedback loop for the parameter values indicated by the cross on the left, and for four different values of $\Omega$. For small system sizes, the stochastic system shows a bimodal stationary probability distribution. With increasing system size, the peak at small concentrations becomes smaller and vanishes above $\Omega_{c}$, where also the second stable fixed point of the convective field vanishes. Note that the bottom two histograms are plotted on a logarithmic scale.}
\label{fig:figure6}
\end{figure}

For this set of reactions, the convective field of the double-positive loop is
\begin{equation}
\renewcommand\arraystretch{1.4}
\vec{\alpha}(x,y) = \begin{pmatrix}b_{x} + m_{x}\frac{y^{n_{y}}}{\theta_{y}^{n_{y}} + y^{n_{y}}} - x\\ b_{y} + m_{y}\frac{\theta_{x}^{n_{x}}}{x^{n_{x}} + x^{n_{x}}} -dy\end{pmatrix} - \frac{1}{2\Omega}\begin{pmatrix} 1 \\ d \end{pmatrix}.
\label{eq:KonvektionsfeldPFLpp}
\end{equation}
We can now derive the stochastic stability diagram of system~\eqref{eq:ReaktionsgleichungenPFLpp} in the same way as before, defining
\begin{equation}
    \beta_{i} = \frac{b_{i}}{\theta_{i}} 
\end{equation}
in addition to~\eqref{eq:VariablenwechselPFLnn}. From the rescaled convective field
\begin{equation}
\renewcommand\arraystretch{1.1}
\vec{\tilde{\alpha}}(\xi,\upsilon) = \begin{pmatrix}\beta_{x} + \mu_{x}\frac{\upsilon^{n_{y}}}{1 + \upsilon^{n_{y}}} - \xi\\ \beta_{y} + \mu_{y}\frac{\xi^{n_{x}}}{1 + \xi^{n_{x}}} -d\upsilon\end{pmatrix} - \begin{pmatrix} \Delta_{x} \\ d\Delta_{y} \end{pmatrix}
\label{eq:KonvektionsfeldPFLpp_rescaled}
\end{equation}
we obtain again a self-consistency equation 
\begin{equation}
    \upsilon^{*} = \tilde{\beta}_{y} + a\frac{\xi^{*}(\upsilon^{*})^{n_{x}}}{1 + \xi^{*}(\upsilon^{*})^{n_{x}}} - \Delta_{y}
    \label{eq:SelbstkonsistenzgleichungPFLpp}
\end{equation}
by setting $\vec{\tilde{\alpha}}(\xi^{*}, \upsilon^{*}) = 0$. Here, 
\begin{equation}
    \tilde{\beta}_{y} = \frac{\beta_{y}}{d}
\end{equation}
is defined similarly to the parameter $a$ in~\eqref{eq:PFLnnDefinition_a}. Figure~\ref{fig:figure6}(a) shows a stability diagram of~\eqref{eq:ReaktionsgleichungenPFLpp}, where the shaded parameter regions indicate that~\eqref{eq:SelbstkonsistenzgleichungPFLpp} has more than 2 solutions, i.e., the convective field shows bistability. Figure~\ref{fig:figure6}(b) shows that the transition to bistability of the convective field is accompanied by a transition  to bimodality in the stationary probability distribution obtained from computer simulations of the reaction network. The stochastic system exhibits two peaks, or at least a peak and a shoulder, for system sizes below $\Omega_{c}$ and transitions to unimodality for $\Omega > \Omega_{c}$. Compared to the examples of the double-negative loop, the two peaks of the stationary probability distribution are much better separated. 

Compared to the double-negative feedback loop, the double-positive feedback loop shows a better pronounced phenomenological bifurcation of the stationary probability distribution with changing system size. This is plausible from the fact that the two peaks of the double-positive feedback loop differ considerably in the total number of molecules, while the total number of molecules is of the same order at the two peaks for the double-negative feedback loop (where the concentration of one protein is large and that of the other small).
Concordantly, the entries of the diffusion matrix are much larger at one fixed point than at the other for the double-positive loop. From the consideration of the one-dimensional feedback loop, we have learned that the change of the strength of diffusion along the line that connects the two maxima is responsible for the shift of the distance between maxima and minima (or saddle points) with changing system size. It is this resemblance of the double-positive feedback loop with a one-dimensional system that we consider responsible for the good agreement between the bifurcation of the convective field and the phenomenological bifurcation of the stationary probability distribution.


\subsection{A two-dimensional oscillating system: the Brusselator}
\label{ssec:Brusselator}
As our last model system,  we choose the Brusselator~\cite{Nicolis1977}, a reaction system of two species capable of oscillations. The Brusselator consists of the following reactions of two species X and Y
\begin{align}
\begin{split}
\ce{\emptyset &<-->[1][1] X} \qquad \ce{X ->[$b$] Y} \qquad \ce{2X + Y ->[$a$] 3X}
\end{split}
\label{eq:ReaktionsgleichungenBrusselator}
\end{align}
with positive reaction rates $a$ and $b$. The macroscopic rate equations of system~\eqref{eq:ReaktionsgleichungenBrusselator} are given by
\begin{equation}
\begin{pmatrix} \dot{x} \\ \dot{y} \end{pmatrix} = \begin{pmatrix}1 - (b+1)x +ax^{2}y \\bx - ax^{2}y\end{pmatrix}
\label{eq:RatengleichungBruesselator}
\end{equation} 
and have the fixed point $(x^{*}, y^{*}) = (1, \frac{b}{a})$. While the steady state of~\eqref{eq:RatengleichungBruesselator} is stable for $b  < (1+a)$, it becomes unstable for $b > (1+a)$. In the latter case, the unstable fixed point is enclosed by a stable limit cycle which emerges from a Hopf bifurcation at $b_{c} = (1+a)$. This is the only bifurcation shown by  the macroscopic rate equations of the Brusselator. 

The convective field of the Brusselator
\begin{equation}
\vec{\alpha}(x,y) = \begin{pmatrix}1 - (b+1)x +ax^{2}y \\bx - ax^{2}y\end{pmatrix} - \frac{1}{2\Omega}\begin{pmatrix} 1 + b - ax^{2} + 2axy \\ -b + ax^{2} + 2axy \end{pmatrix}
\label{eq:KonvektionsfeldBruesselator}
\end{equation} 
also has one fixed point. Like the macroscopic model, the concective field can undergo a Hopf bifurcation. The bifurcation lines of the stochastic stability diagrams can be derived from the condition that the trace of the Jacobian of $\vec{\alpha}(x,y)$ at the fixed point must vanish~\cite{Strogatz1994}, i.e., $\partial_x\alpha_x+\partial_y\alpha_y=0$. Besides the Hopf bifurcation, the convective field~\eqref{eq:KonvektionsfeldBruesselator} does not undergo any other bifurcations. 

\begin{figure}[!htbp]
\begin{center}
\includegraphics[width=0.6\columnwidth]{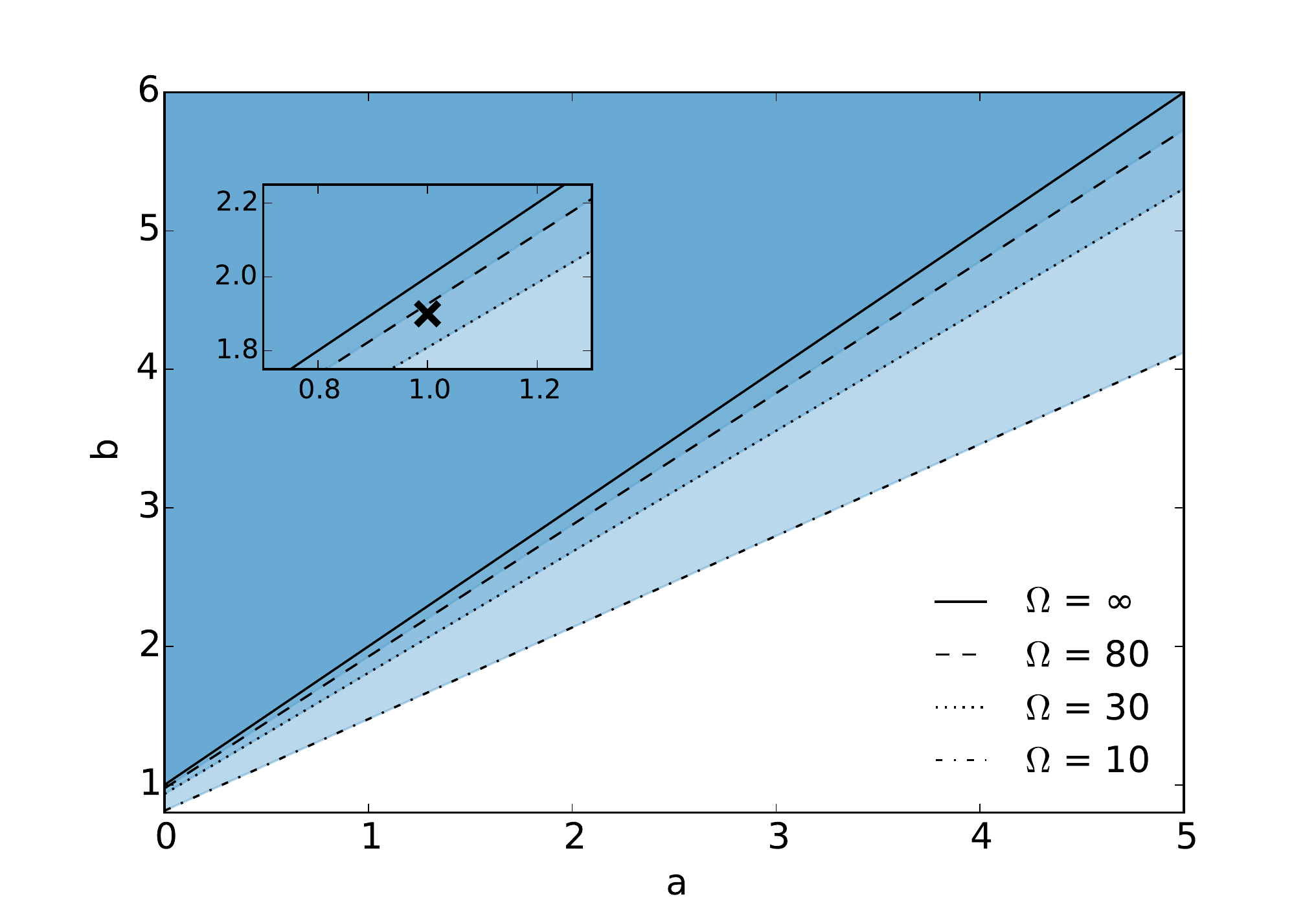}
\caption{Stability diagram of the Brusselator for various system sizes $\Omega$. Along the lines, the convective field undergoes a Hopf bifurcation. Above the lines, the dynamical systems governed by the convective field show a stable limit cycle. The corresponding regions are shaded. The uppermost shaded region corresponds to the region of stable limit cycles of the macroscopic system. Inset: Parameter values of the systems shown in figure~\ref{fig:figure8}.}
\label{fig:figure7}
\end{center}
\end{figure}

The stability diagram of the convective field of the Brusselator is shown in figure~\ref{fig:figure7} for different system sizes $\Omega$. The lines separate parameter regions with oscillatory and non-oscillatory behavior, respectively. Above the phase boundary, the dynamics given by the convective field show a stable limit cycle. Below the phase boundary, only a stable fixed point exists. The figure shows that the convective field shows a limit cycle for a larger proportion of parameter space than the macroscopic system does. This means that with decreasing system size the convective field can undergo a Hopf bifurcation that leads to a limit cycle.

\begin{figure}[!htbp]
\centering
\subfigure{\includegraphics[height=0.26\textheight]{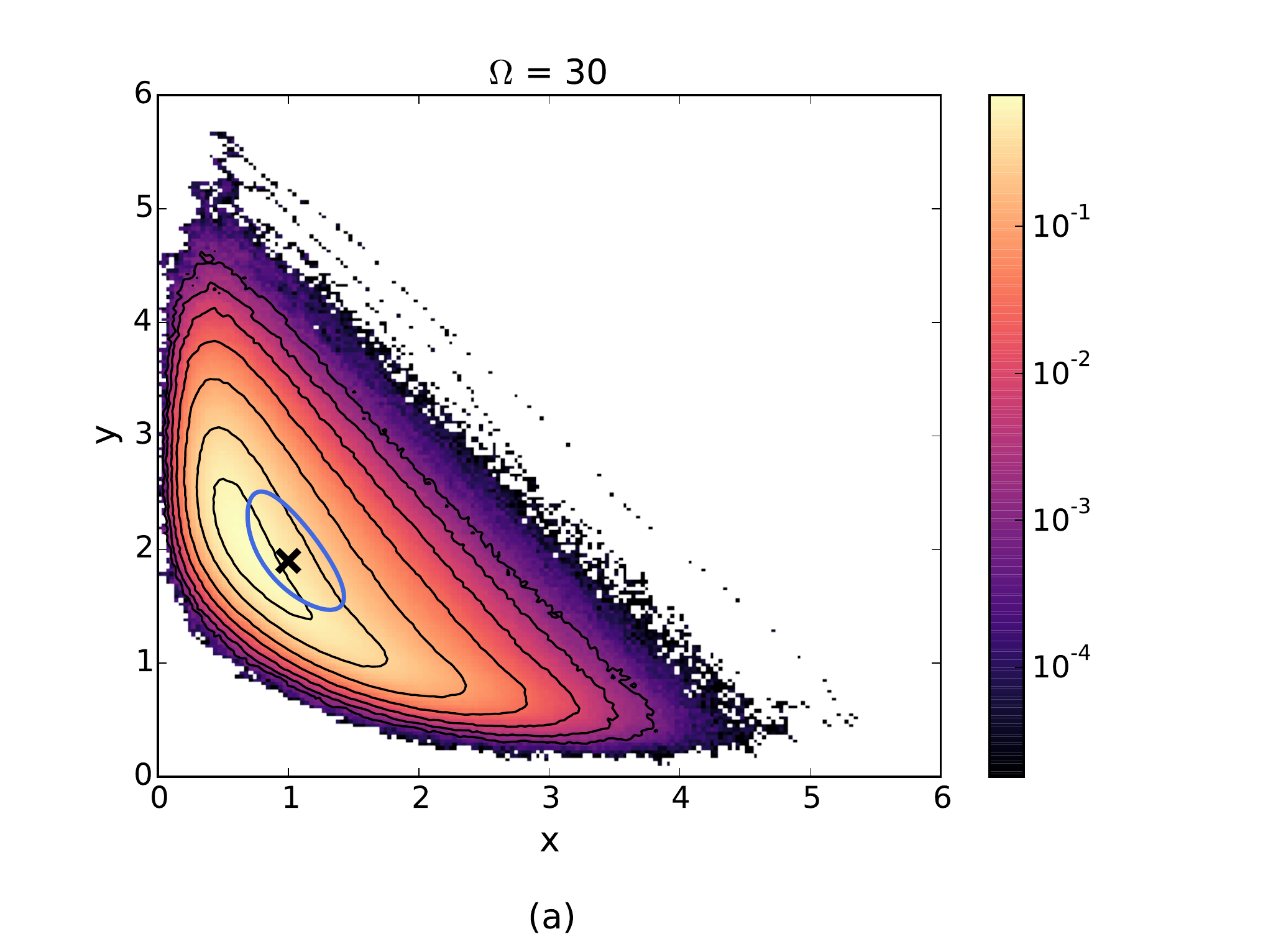}}
\subfigure{\includegraphics[height=0.26\textheight]{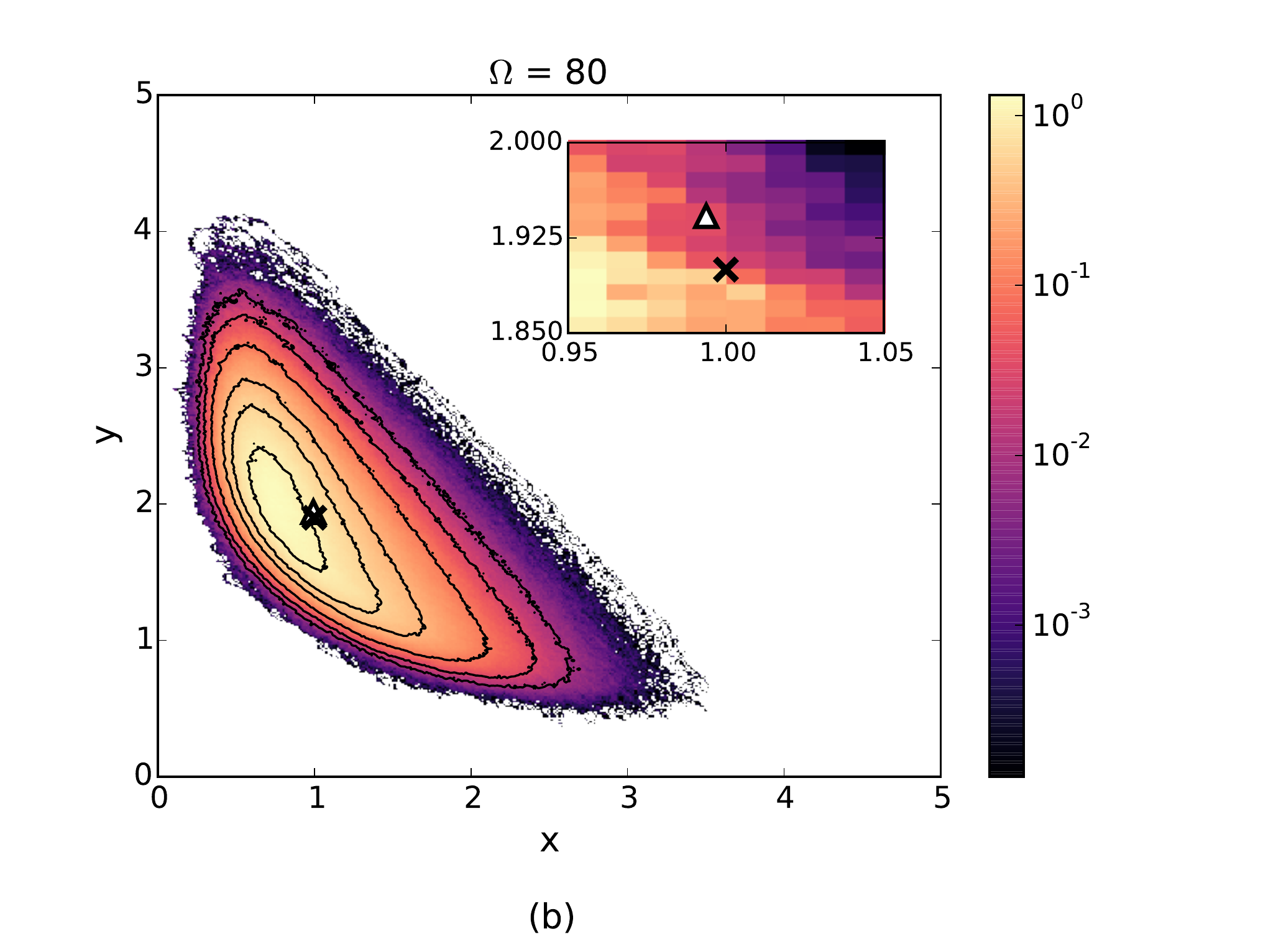}}
\caption{Stationary probability distributions obtained from stochastic simulations of the Brusselator~\eqref{eq:ReaktionsgleichungenBrusselator} for $a = 1$, $b = 1.9$ and (a) $\Omega = 30$ or (b) $80$. The stable steady state of the macroscopic system is marked by a black cross in both figures. The limit cycle of the convective field is drawn in blue for $\Omega = 30$, its stable fixed point for $\Omega = 80$ is marked by a white triangle. The stable fixed points of $\vec{\alpha}(x,y)$ and the macroscopic system~\eqref{eq:RatengleichungBruesselator} lie very close together for $\Omega = 80$, as shown by the inset in figure~\ref{fig:figure8}(b). Whilst the convective field bifurcates under variation of $\Omega$, the simulated stationary probability distributions do not differ qualitatively. The distributions were obtained from $5\cdot10^{7}$ data points each.} 
\label{fig:figure8}
\end{figure}

Figure~\ref{fig:figure8} shows that the stationary probability distribution remains unimodal when the system size is lowered to a value where the convective field has undergone the Hopf bifurcation.

\begin{figure}[!htbp]
\centering
\subfigure{\includegraphics[width=0.64\columnwidth]{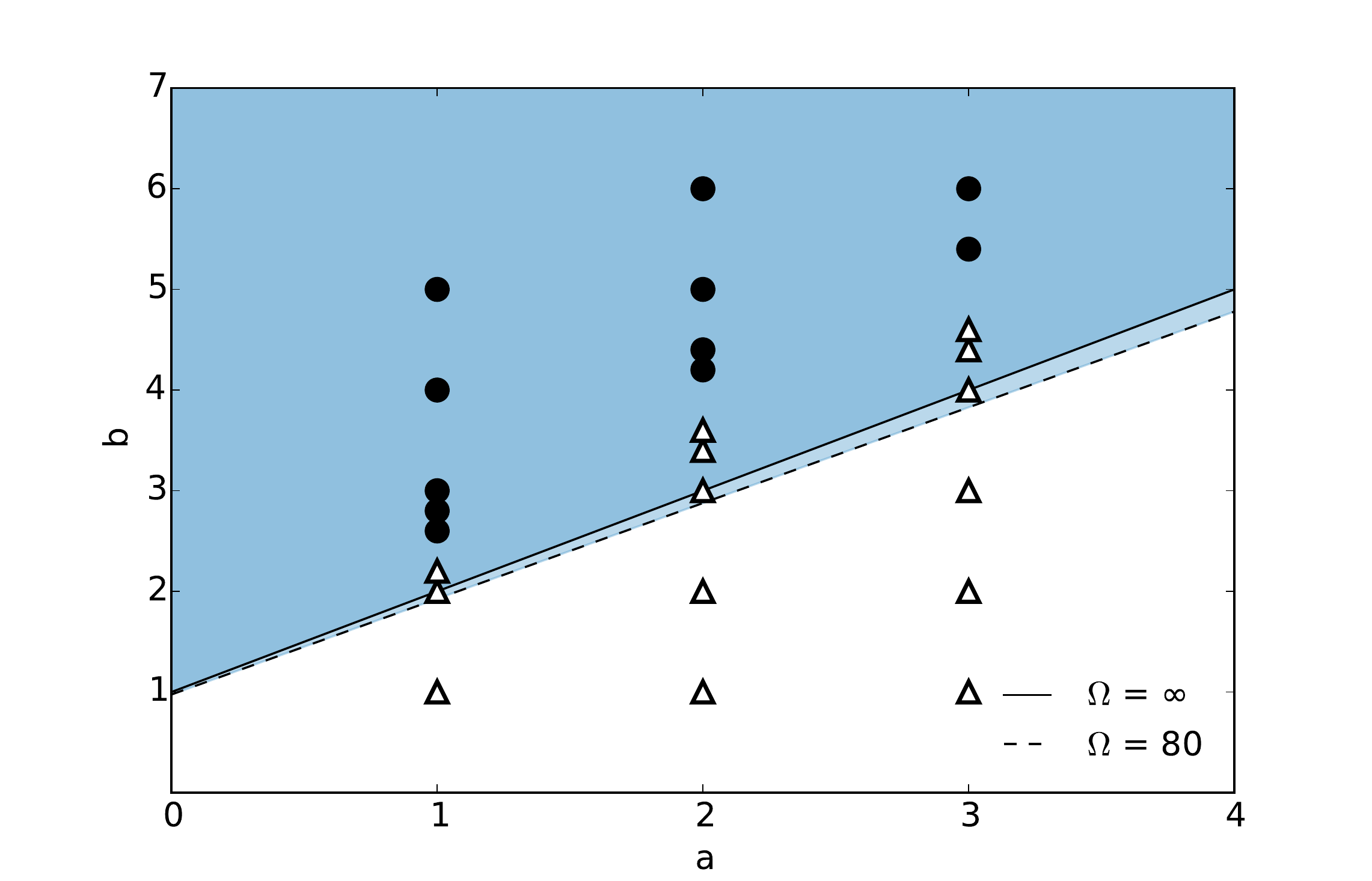}}
\caption{Comparison between the dynamics of $\vec{\alpha}(x,y)$ and the macroscopic system and the shape of the stationary probability distribution of the Brusselator. For white triangles, the stationary probability distribution is not crater-shaped, for black circles it is.  Whenever the (non-)existence of a crater of the stationary probability distributions was not apparent by eye, the contour lines were computed and checked for signs of a crater, similar to the procedure shown in figure~\ref{fig:figure8}. Shaded regions indicate the existence of limit cycles of $\vec \alpha(x,y)$ and $\vec f(x,y)$ respectively.}
\label{fig:figure9}
\end{figure}

Figure~\ref{fig:figure9} compares the attractors of the convective field to the shape of the stationary probability distribution obtained from computer simulations. As one can see, the stationary probability distribution makes the transition to the crater shape only well within the regime where the macroscopic system oscillates. So while for the convective field stable limit cycles can emerge under a decrease of the system size, the picture derived from simulations is the opposite: Here, a decrease of $\Omega$ can actually turn the crater-shaped stationary probability distribution of the macroscopic system into a unimodal one.

This result shows that the correspondence between system-size induced phenomenological bifurcations of the stationary probability distribution and bifurcations of the convective field breaks down for the Hopf bifurcation, which cannot occur in one dimension for which the idea was formulated. Indeed, the finding that a decrease of system size turns a crater into a peak close to a Hopf bifurcation can be made plausible: Exactly at the Hopf bifurcation of the convective field $\vec \alpha(x,y)$, the vector field $\vec \alpha(x,y)$ shows elliptic trajectories around the marginally stable fixed point. This means that any flow of probability in the direction perpendicular to $\vec \alpha(x,y)$ is due to undirected diffusion (because directed diffusion is already included in $\vec \alpha(x,y)$). However, undirected diffusion moves the same amount of probability to the outside and to the inside of each closed trajectory of $\vec\alpha(x,y)$. Since the area between ellipses decreases towards the center of the ellipses, probability accumulates there, leading to a peak of the stationary probability distribution. This means that at the point where the convective field undergoes a Hopf bifurcation the stationary probability distribution still shows a local maximum. This effect becomes stronger with increasing importance of the diffusion matrix, i.e., with decreasing system size.

\section{Discussion}
\label{sec:Discussion}
In this paper we investigated the relation between phenomenological bifurcations of the stationary distribution of chemical reaction networks and bifurcations of the convective field. We focused on parameter regions where a change in system size induces a bifurcation in the convective field. 

For one-dimensional systems, these two types of bifurcations coincide in parameter space, and we exploited this fact to find regions in parameter space where a positive autoregulator undergoes a phenomenological saddle-node bifurcation.  Since the position of the bifurcation lines in parameter space changes with system size or, equivalently, with the discreteness parameter~\cite{Scott2007}, a change of the discreteness parameter can induce a bifurcation. We showed that these system size-dependent p-bifurcations trace back to the concentration dependence of the diffusion coefficient, the effects of which are included in the convective field.

We studied the relation between saddle-node bifurcations of the convective field and of the stationary probability distribution also for two-dimensional systems, using the example of a two-species positive feedback loop. For the double-positive feedback loop, we found a very good agreement between the system-size dependent double-peak structure of the stationary probability distribution and a change in the number of sinks of the convective field. For the double-negative feedback loop, we found that in the vicinity of the saddle-node bifurcation of the convective field the stationary probability distribution showed a peak and a shoulder instead of two peaks.  
In situations where the stationary probability distribution shows two peaks for large system sizes, we found that their weights can be vastly different, so that the small peak may not really be relevant for the stochastic dynamics. This means that the correspondence between saddle-node bifurcations of the convective field and p-saddle-node bifurcations of the stationary probability distribution is less good for the double-negative feedback loop. Nevertheless, the shoulder indicates that closeness to a saddle-node bifurcation of the convective field implies closeness of the reaction network to a p-saddle-node bifurcation.  

In our study, we did not evaluate quantitatively the relative weight of the two peaks. 
Endres~\cite{Endres2015} argued that with increasing system size one of the two modes becomes increasingly favored as switching events become increasingly rare. This fits together with the trend visible in our figure \ref{fig:figure4}(b) that the heights of the two peaks become more different with increasing system size $\Omega$.   

For the Brusselator, we found that system-size induced Hopf bifurcations of the convective field were not associated with phenomenological bifurcations of stationary probability distribution. Limit cycles of the convective field did not correspond to circular ridges of stationary probability distributions for parameter sets belonging to the macroscopically non-oscillatory regime. Moreover, while decreasing system size induced a Hopf bifurcation in the convective field, the stationary probability distribution developed a crater-like shape only sufficiently deep in the macroscopically oscillatory regime. By lowering the system size and thus the molecule number we could transform a crater-shaped stationary probability distribution into a unimodal one. Just recently, Constantino and Kaznessis pointed out this effect, arguing that it represents a new kind of bifurcation unknown in the macroscopic limit~\cite{Constantino2019}. These results do, however, not rule out the existence of a characteristic frequency in the stochastic system. Indeed, the power spectrum can show such a  characteristic frequency even when the macroscopic model has a stable fixed point~\cite{McKane2005}. 

Taking all these results together, we observe that the correspondence between bifurcations of the convective field and of the stationary probability distribution is better when the bifurcation is more similar to that of a one-dimensional system: When the considered model is one-dimensional, the correspondence is of course perfect as stationary currents are exactly zero. The two-dimensional model that is closest to the one-dimensional case, the double-positive feedback loop, shows also a good correspondence between the phenomenological bifurcation of the stationary probability distribution and the bifurcation of the convective field. For the double-negative feedback loop, there is no longer a continuous increase in the size of the entries of the diffusion matrix as one moves from one stable fixed point to the other. The analogy with the one-dimensional system is therefore less clear, and the system-size induced bifurcation of the convective field is reflected in the stationary probability distribution not as clearly as for the double-positive loop. For the Brusselator, which undergoes a Hopf bifurcation, there is no analogy at all in one-dimensional systems. Even in the limit of infinite system size the stationary current does not vanish, and there is therefore no limit in which this current is small and the correspondence between the two types of bifurcations good. 

Since stationary currents usually do not vanish even for saddle-node bifurcations ~\cite{Ceccato2018}, there is a need for investigating these stationary currents in order to better understand the factors influencing phenomenological bifurcations. 


\section*{Acknowledgements}

We thank Johannes Falk for helpful discussion.
This work was supported in part by the Landes-Offensive zur Entwicklung wissenschaftlich-ökonomischer Exzellenz (LOEWE; initiative to increase research excellence in the state of Hessen, Germany) within the LOEWE-Schwerpunkt CompuGene.
We acknowledge support by the Open Access Publishing Fund of Technische Universität Darmstadt. 



\end{document}